\documentclass[fleqn,twoside]{article}
\usepackage{amsfonts}
\usepackage{latexsym}
\usepackage{gc}

\heads
{K.A. Bronnikov and S.V. Grinyok}
{Conformal continuations and wormhole instability in scalar-tensor gravity}

\def\mn{_{\mu\nu}}
\def\MN{^{\mu\nu}}
\def\mN{_\mu^\nu}

\def\og{{\overline g}}

\def\M{{\mathbb{M}}}
\def\N{{\mathbb{N}}}
\def\R{{\mathbb R}}
\def\S{{\mathbb S}}
\def\oM{{\overline \M}}
\def\cR{{\cal R}}

\def\Str{\mbox{$\S_{\rm trans}$}}
\def\ME{\mbox{$\M_{\rm E}$}}
\def\MJ{\mbox{$\M_{\rm J}$}}

\def\GR{general relativity}
\def\sph{spherically symmetric}
\def\ssph{static, spherically symmetric}

\def\wh{wormhole}
\def\whs{wormholes}

\begin{document}
\twocolumn[
\prepno{gr-qc/0411063}{\GC {10} 237 (2004)}

\bigskip

\Title {Conformal continuations and wormhole instability\yy
    in scalar-tensor gravity}

\Authors
{K.A. Bronnikov\foom 1} {and S.V. Grinyok\foom 2}
{Centre for Gravitation and Fundam. Metrology, VNIIMS,
        3-1 M. Ulyanovoy St., Moscow 119313, Russia;\\
 Institute of Gravitation and Cosmology,
        PFUR, 6 Miklukho-Maklaya St., Moscow 117198, Russia}
{Bauman Moscow State Technical University,  Moscow, Russia}

\Abstract
{We study the stability of static, spherically symmetric, traversable
wormholes existing due to conformal continuations in a class of
scalar-tensor theories with zero scalar field potential (so that Fisher's
well-known scalar-vacuum solution holds in the Einstein conformal frame).
Specific examples of such wormholes are those with  nonminimally 
(e.g., conformally) coupled scalar fields. 
All boundary conditions for scalar and metric perturbations are taken into
account. All such wormholes are shown to be unstable under spherically
symmetric perturbations. The instability is proved analytically with the aid
of the theory of self-adjoint operators in Hilbert space and is confirmed by
a numerical computation.}

] 
\email 1 {kb20@yandex.ru}
\email 2 {stepanv1@mail.ru}

\section{Introduction}

   In our recent paper \cite{bg01} we have considered \sph\ perturbations
   of \wh\ solutions to the Einstein-massless scalar field equations which
   exist for scalar fields nonminimally coupled to gravity
   \cite{br73,bar-vis99}. The equations of motion were reduced to a
   single wave equation for the scalar field perturbation which in this
   case comprises the only dynamical degree of freedom. An analysis of this
   wave equation leads to the conclusion that such \whs\ are unstable under
   \sph\ (monopole) perturbations, and this instabilty is of catastrophic
   nature since the increment of perturbation growth has no upper bound.

   In this paper we continue this study and extend it in two respects.
   First, we discuss more general background configurations, namely, \ssph\
   \whs\ which appear in arbitrary scalar-tensor theories (STT) of gravity
   in which the effective gravitational constant can change its sign due to
   conformal continuation (CC) \cite{br02-CC}. The investigation is,
   however, restricted to massless fields for which Fisher's well-known
   solution holds in the Einstein frame. Second, we examine the problem
   in more detail, including the behaviour of metric perturbations related
   to those of the scalar field. A physically meaningful metric perturbation
   of an initially regular configuration should be regular everywhere. This
   requirement turns out to impose an additional constraint on the scalar
   field perturbations, which makes the stability problem quite nontrivial.
   We finally prove that there exists at least a single growing mode of
   physically meaningful perturbations, i.e., such \whs\ are indeed
   unstable. However, contrary to the conclusion of Ref.\,\cite{bg01}, the
   perturbation grows at a finite rate.

   We thus find that gravitational instabilities, whose existence seems
   to be quite natural at surfaces where the gravitational coupling changes
   its sign (see, e.g., Ref.\,\cite{star81} for a discussion in a
   cosmological setting), still need much effort in their detailed study and
   even discovery.

   The paper is organized as follows. \sect 2 is a brief description of the
   background static configuration and its place among more general
   configurations of this kind, i.e., \ssph\ \wh\ solutions of a general
   class of scalar-tensor theories (STT) admitting conformal continuations
   (CCs) \cite{br02-CC}. \sect 3 discusses \sph\ perturbation equations and
   the corresponding gauge freedom. \sect 4 is devoted to a stability
   investigation for \whs, both analytical, using the theory of self-adjoint
   operators in Hilbert space, and numerical.

\section{Conformal continuations and wormhole solutions
         of scalar-tensor theories}

\subsection {STT in Jordan and Einstein pictures}

   Consider the general (Bergmann-Wagoner-Nordtvedt) class of STT, where
   gravity is characterized by the metric $g\mn$ and the scalar field
   $\phi$; the action is
\bearr
    S = \int d^4 x \sqrt{g}\big\{ f(\phi) \cR [g]           \label{act}
                + h(\phi)g\MN\phi_{,\mu}\phi_{,\nu}
\nnn \inch\cm
                    - 2 U(\phi) + 16\pi G\,L_m \big\}.
\ear
   Here $\cR [g]$ is the scalar curvature, $f,\ h$ and $U$ are certain
   functions of $\phi$, varying from theory to theory, $L_m$ is the matter
   Lagrangian, and $G$ is the gravitational constant, not necessarily
   coinciding with its Newtonian value.

   The action (\ref{act}) is simplified by the
   well-known conformal mapping \cite{Wagoner1970}
\beq
    g\mn = \og\mn/|f(\phi)|,                             \label{map}
\eeq
   accompanied by the scalar field transformation $\phi\mapsto \psi$ such
   that
\beq                                                        \label{ps-f}
   \frac{d\psi}{d\phi}= \pm \frac{\sqrt{|l(\phi)|}}{f(\phi)},
      \qquad    l(\phi) \eqdef fh +\frac 32
                                 \biggl(\frac{df}{d\phi}\biggr)^2.
\eeq
   In terms of $\og\mn$ and $\psi$, the action for $U = L_m = 0$, the case
   of massless scalar-vacuum fields to be considered here, takes the form
\beq                   \nq                                    \label{act-E}
    S = \int d^4 x \sqrt{\og} (\sign f)
      \big\{ \cR [\og] + [\sign l(\phi)]\og \MN \psi_{,\mu} \psi_{\nu}\big\}
\eeq
   (up to a boundary term which does not affect the field equations).
   Here $R[\og]$ is the Ricci scalar obtained from $\og\mn$.

   The space-time $\MJ = \M [g]$ with the metric $g\mn$ is referred to as the
   {\it Jordan conformal frame\/} (or picture), generally regarded as the
   physical frame in STT; the {\it Einstein conformal frame\/} $\ME=\oM[\og]$
   with the field $\psi$ then plays an auxiliary role (see, however,
   discussions of the physical meaning of various conformal frames in
   \cite{faraoni,bm-eric} and references therein). The action (\ref{act-E})
   corresponds to conventional \GR\ if $f>0$, and the normal sign of scalar
   kinetic energy is obtained for $l(\phi) > 0$. Scalar fields in anomalous
   STT, in which $l(\phi) <0$, lead to a kinetic term in (\ref{act-E}) with
   a ``wrong'' sign, are called phantom scalar fields. Such fields (with
   different potentials) are sometimes invoked in modern cosmological
   studies to describe dark energy.

   Exact \ssph\ scalar-vacuum solutions of the theory (\ref{act}) are well
   known \cite{fish,br73}. Among them, \wh\ solutions are generic in the
   case of phantom scalar fields \cite{br73}. Their stress-energy tensor
   $T\mn$ manifestly violates the null energy condition (NEC) $T\mn
   k^{\mu}k^{\nu} \geq 0$, $k_\mu k^\mu =0$, such violation being a
   necessary condition for \wh\ existence \cite{hoh-vis}, therefore \wh\
   solutions in their presence would have been naturally expected. One can
   note that, according to such solutions, both space-times $\MJ$ and $\ME$
   have \wh\ properties, i.e., represent regular static traversable bridges
   between two flat asymptotics. The stability of such configurations has
   also been proved \cite{cold,picon02} by a direct study of perturbation
   equations, though it seems quite strange for a field system with energy
   density unbounded from below. This question evidently deserves further
   investigation.

   Our interest here will be in other \wh\ solutions which appear in STT
   with $l(\phi) >0$, in cases when the space-time manifold $\ME$ is mapped,
   according to (\ref{map}), to only a part of $\MJ$; this phenomenon
   was named {\it conformal continuation\/} (CC) \cite{br02-CC,br-ustron}.
   The Jordan space-time $\MJ$ is then globally regular, represents a \wh,
   and it two non-intersecting parts map to two singular space-times
   $\ME$ and $\ME'$.

\subsection{Fisher's solution and its conformal continuations}

   The general \ssph\ solution to the Einstein-scalar equations that follow
   from (\ref{act-E}), was first found by Fisher \cite{fish} and was
   repeatedly rediscovered afterwards. Let us write it in the form suggested
   in \cite{br73}, restricting ourselves to the ``normal'' case $l > 0$:
\bear                                                      \label{psi}
    \psi(u) \eql Cu + \psi_0,
\\                                                         \label{ds-E}
   ds_{\rm E}^2 \eql
        \e^{2\gamma(u)}dt^2 -\e^{2\alpha(u)}du^2-\e^{2\beta(u)}d\Omega^2
\nnn\nq
    = \e^{-2mu} dt^2-\frac{k^2\e^{2mu}}{\sinh^2(ku)}
                      \left[\frac{k^2du^2}{\sinh^2(ku)} + d\Omega^2\right]
\ear
   where the subscript ``E'' stands for the Einstein frame; $d\Omega^2=
   d\theta^2 + \sin^2\theta\,d\phi^2$ is the linear element on a unit sphere;
   $C$ (scalar charge), $m >0$ (mass in geometric units), $k>0$ and $\psi_0$
   are integration constants, of which the first three are related by
\beq                                                      \label{condKCH}
        k^2 = m^2 + \half C^2.
\eeq
   Without loss of generality we put $C > 0$ and $\psi_0{=}0$. We are here
   using the harmonic radial coordinate $u\in \R_+$ in $\ME[\og]$,
   satisfying the coordinate condition $\alpha = 2\beta + \gamma$.

   Another convenient form of the solution is obtained in isotropic
   coordinates: with $y = \tanh (ku/2)$, \eqs (\ref{psi}), (\ref{ds-E})
   are converted to
\bear
    \psi (y) \eql\frac{C}{k}\ln \biggl|\frac{1+y}{1-y}\biggr|,  \label{psi_}
\\
    ds^2_{\rm E} \eql                                          \label{ds-E_}
            A(y)\,dt^2
            - \frac{k^2(1-y^2)^2}{y^4 A(y)} (dy^2 + y^2 d\Omega^2),
\nn
     A(y) \eql \biggl|\frac{1-y}{1+y}\biggr|^{2m/k}.
\ear

   The solution is asymptotically flat at $u\to 0$ ($y \to 0$), has
   no horizon when $C\ne 0$ and is singular at the centre ($u\to \infty$, $y
   \to 1-0$, $\psi \to \infty$). When the scalar field is ``switched off''
   ($C=0$, $k=m$), the Schwarzschild solution is recovered.

   A feature of importance is the invariance of (\ref{psi_}), (\ref{ds-E_})
   under the inversion $y \mapsto 1/y$, noticed probably for the first
   time by Mitskievich \cite {mitz}. Due to this invariance, the solution
   (\ref{psi_}), (\ref{ds-E_}) considered in the range $y > 1$ describes
   quite a similar configuration, but now $y\to \infty$ is a flat asymptotic
   and $y\to 1+0$ is a singular centre. An attempt to unify the two ranges
   of $y$, or, in other words, the two copies of Fisher's solution, is
   meaningless due to the singularity at $y = 1$. We shall see that such a
   unification, leading to a wormhole, is achieved in $\MJ[g]$\ where the
   singularity is smoothed out (in case $C = \sqrt{6}m$) owing to the
   conformal factor.

   The corresponding Jordan-frame solution for any $f(\phi)$ and $h(\phi)$
   such that $l(\phi) > 0$ are obtained from (\ref{psi}), (\ref{ds-E}) using
   (\ref{map}), (\ref{ps-f}). If the function $f(\phi)$ is everywhere
   finite, $\MJ$ has the same basic properties as $\ME$.

   However, according to \cite{br02-CC}, there is a class of STT in which
   some solutions produce structures of $\MJ$ drastically different from that
   of $\ME$. Namely, let us use the $\phi$ field reparametrization freedom
   of the action (\ref{act}) [$\phi = \phi(\phi_{\rm new})$] and fix the
   parametrization by putting in (\ref{act}) $h(\phi) \equiv 1$. Then
   \cite{br02-CC}, if the function $f(\phi)$ has a simple zero at some $\phi
   = \phi_0$, there is a subfamily of \ssph\ solutions to the field
   equations admitting a CC. The latter means that a singular surface in
   $\ME$, corresponding to $\phi=\phi_0$, maps according to (\ref{map}) to a
   regular surface \Str\ in $\MJ$. Then $\M$ can be continued in a regular
   manner through this surface, and the global properties of $\MJ$ can be
   considerably richer than those of $\ME$: in the new region one can
   possibly find, e.g., a horizon or another spatial infinity. The above
   result was obtained in \cite{br02-CC} for STT (\ref{act}) in arbitrary
   dimensions and with arbitrary potentials $U(\phi)$. It was also shown
   \cite{br02-CC} that a \wh\ was a generic type of a conformally continued
   Jordan-frame manifold. Before studying perturbations of such generic
   solutions (but with $U(\phi) \equiv 0$, so that we have Fisher's solution
   in the Einstein picture), we discuss a specific example which makes
   evident the relations between Einstein and Jordan quantities.

\subsection {Example: wormholes with a conformally coupled scalar field}

    A particular example of a CC is given by a free massless conformally
    coupled scalar field in GR. The latter is obtained when we put in
    (\ref{act})
\beq  \nq\,                                                    \label{conf}
     f(\phi) = 1 - \phi^2/6, \quad\  h(\phi) \equiv 1.
                             \quad\  U(\phi) = L_m =0.
\eeq
    A transition sphere \Str, if any, corresponds to $\phi^2 {=} 6$.

    The transformation (\ref{ps-f}) now takes the form
\beq                                                    \label{trans-f}
    \frac{d\psi}{d\phi} = \frac {1}{1 - \phi^2/6}.
\eeq
    We assume that spatial infinity, where $\psi\to 0$, corresponds in the
    Jordan space-time $\MJ$ to $|\phi| < \sqrt{6}$, where $f(\phi) > 0$, so
    that the gravitational coupling has its normal sign. Then
    (\ref{trans-f}) gives
\beq                                             \label{psi3}
     \psi = \sqrt{6}\, \tanh^{-1} (\phi/\sqrt{6}) + \psi_0,
            \qquad \psi_0 = \const.
\eeq

    Using (\ref{psi}) and (\ref{ds-E}), it is now easy to write the metric
    in the Jordan picture.

    A CC through the sphere \Str\ ($u=\infty$, $y=1$, $\phi=\sqrt{6}$),
    which is singular in $\ME$, is obtained when the infinity of the
    conformal factor $1/f$ in (\ref{map}) compensates the zero of both
    $\og_{tt}$ and $\og_{\theta\theta}$ simultaneously.  This happens when,
    in accord with (\ref{condKCH}),
\beq                                                        \label{k=2h}
        k = 2m =2C/\sqrt{6},
\eeq
    which selects a special subfamily among all solutions. We will
    restrict the consideration to this subfamily.

    In terms of the isotropic coordinate $y$,
    the solution in the Jordan picture has the form \cite{br73}
\bearr                                   \nq \,
     ds^2 = \frac{(1{+}yy_0)^2}{1-y_0^2} \biggl[ \frac{dt^2}{(1{+}y)^2}
            - \frac{m^2(1{+}y)^2}{y^4} (dy^2 + y^2 d\Omega^2) \biggr],
\nnnv
     \phi (y) = \sqrt{6} \frac{y+y_0}{1 + yy_0},          \label{con-y}
\ear
    where $y_0 = \tanh (\psi_0/\sqrt{6})$. The range $0 < y < 1$,
    describing the whole manifold \ME\ in the Fisher solution, corresponds
    to only a region $\MJ'$ of the manifold \MJ\ of the solution
    (\ref{con-y}). In all cases, $y=0$ corresponds to a flat asymptotic,
    where $\phi \to \sqrt{6}y_0 < \sqrt{6}$. The global properties of the
    solution depend on the sign of $y_0$:

\medskip\noi
    {\bf a)} $y_0 < 0$. The solution is defined in the range $0 < y
    <1/|y_0|$.  At $y = 1/|y_0|$, there is a naked attracting central
    singularity:  $g_{tt}\to 0$, $r^2\to 0$, $\phi\to\infty$.

\medskip\noi
{\bf b)} $y_0=0$, $\phi=\sqrt{6}y$, $y\in \R_+$. In this case it is helpful
    to pass to the conventional radial coordinate $r$, substituting $y =
    m/(r-m)$. The solution
\bear
     ds^2 \eql (1-m/r)^2{dt^2} - \frac{dr^2}{(1-m/r)^2} -r^2 d\Omega^2,
\nn
     \phi \eql \sqrt{6}m/(r-m)                              \label{con-bh}
\ear
    is the well-known BH with a conformal scalar field \cite{bbm70,bek74}.
    The infinite value of $\phi$ at the horizon $r=m$ does not make the
    metric singular since, as is easily verified, the energy-momentum tensor
    remains finite there. This solution has been shown to be unstable under
    radial perturbations \cite{bk78}.

\medskip\noi
{\bf c)} $y_0 > 0$. This is the {\sl wormhole solution\/} discussed in
    \cite{bg01,br73} and, among other solutions, re-analyzed now. The
    solution is defined in the range $y\in \R_+$.  At $y\to\infty$, we find
    another flat spatial infinity, where $\phi\to \sqrt{6}/y_0$,
    $r^2\to\infty$ and $g_{tt}$ tends to a finite limit.

    The whole manifold \MJ\ can be represented as the union
\beq
    \MJ = \MJ_1 \cup \Str \cup \MJ_2
\eeq
    where the region $\MJ_1$ ($y<1$) is, according to (\ref{map}), in
    one-to-one correspondence with the whole manifold \ME\ of Fisher's
    solution (\ref{psi}), (\ref{ds-E}).  The ``antigravitational''
    ($f(\phi) < 0$) region $\MJ_2$ ($y > 1$) is in a similar correspondence
    with another ``copy'' of Fisher's solution, $\ME'[\og]$, where,
    instead of (\ref{psi3}),
\beq
     \psi = \sqrt{6}\, \coth^{-1} (\phi/\sqrt{6}) + \psi'_0,
                \qquad \psi'_0 = \const.
\eeq
    The metric $\og\mn$ of this second Einstein-frame manifold
    $\ME'$ should also be regularized by the factor $1/f$ on \Str,
    hence the integration constants in it should satisfy the condition
    (\ref{k=2h}). Moreover, one can verify that, to provide a smooth
    transition in the Jordan-frame metric $g\mn$ through \Str, all
    the constants $k$, $h$, $C$ and $\psi_0$ should coincide in $\ME$ and
    $\ME'$. The latter statement is proved using the coordinate $y$ which is
    common on both sides of \Str.

    This example well illustrates the general properties of conformally
    continued solutions \cite{br02-CC}. Namely, in the region beyond \Str,
    there can be a singularity due to $l(\phi) = 0$, as happens in the above
    case a) at $y = 1/|y_0| > 1$. If there is no such singularity, we obtain
    a \wh. Case b), with a horizon, is exceptional, inherent only to the
    field (\ref{conf}) in four dimensions. Thus, for a more general
    action discussed in \cite{bg01,bar-vis00}, with
\beq                                                           \label{L-xi}
     f(\phi) = 1 - \xi\phi^2, \quad\ h(\phi) \equiv 1.
        \quad\  U(\phi) = L_m =0,
\eeq
    with the coupling constant $\xi > 0$, in case $\xi> 1/6$ all solutions
    exhibiting a CC describe \whs, whereas for $\xi < 1/6$ everything
    depends on an integration constant similar to $y_0$ in the above
    example, and we may have either a \wh\ or a naked singularity.

    The stability analysis developed below covers \wh\ solutions obtained
    in the theory (\ref{act}) under the conditions
\bear                                                          \label{condi}
        h\equiv 1, \cm U = L_m = 0, \cm l(\phi) > 0,
\ear
    with an {\it arbitrary\/} function $f(\phi)$, having a simple zero at
    some $\phi=\phi_0$. In other words, we consider massless scalar fields
    in a general non-phantom STT, for which \wh\ solutions exist due to a
    CC.

\section {Spherically symmetric perturbations and gauge freedom}

    Consider small \sph\ (monopole) perturbations of any \ssph\ solution
    of the theory (\ref{act}), (\ref{condi}). The only dynamical degree of
    freedom is evidently related to the scalar field due to the generalized
    Birkhoff theorem \cite{birk}: if we take a time-independent scalar
    field, the equations of motion automatically lead to a static solution.

    We will use, for simplicity, the Einstein conformal frame, since the
    perturbation equations in $\MJ$, being equivalent to those in \ME,
    look much more complicated, and it is even hard to decouple different
    components of the Einstein equations. However, the boundary conditions
    that select physically meaningful perturbation should be formulated for
    variables specified in \MJ\ and only then converted to Einstein-frame
    quantities.

    In the Einstein picture, the equations of motion are the Einstein-scalar
    field equations due to (\ref{act-E})
\bear                                              \label{SE}
    \nabla_\alpha \nabla^\alpha \psi \eql 0,
\\
         R\mN \eql -\psi_{,\mu} \psi^{,\nu}.              \label{EE-E}
\ear

    We now write the metric in \ME\ in the form
\beq                                                       \label{ds}
     ds_{\rm E}^2 =\e^{2\gamma}dt^2 -\e^{2\alpha}du^2-\e^{2\beta}d\Omega^2,
\eeq
    where the functions $\alpha,\ \beta,\ \gamma$ as well as the scalar
    field $\psi$ are split into a static background part and a small (linear)
    time-dependent perturbation:
\[
     \alpha = \alpha(u) + \delta\alpha(u,t)
\]
    where $u$ is a radial coordinate, and similarly for $\beta$, $\gamma$
    and $\psi$.  Now, in addition to the freedom of choosing the radial
    coordinate $u$, we have an additional freedom of specifying the frame of
    reference in perturbed space-time, called {\it gauge freedom\/}. The
    latter makes it possible to specify (by hand) some linear relation
    between the perturbations. Certain care is needed to ensure that the
    resulting perturbation will not be a ``pure gauge'', i.e., will not be
    removable by coordinate transformations.

    Consider an infinitesimal coordinate transformation of the static metric
    (\ref{ds}) preserving its spherical symmetry, i.e.,
\bear                                                        \label{dx^mu}
    x^{\mu}_{\rm new} = x^{\mu}_{\rm old} + \zeta^{\mu}, \cm
    \zeta^\mu = (\bar\eta,\ \bar\xi,\ 0,\ 0),
\ear
    where the time dependence of the perturbations $\bar\xi,\ \bar\eta$ is
    separated:  $\bar\eta = \eta(u) \e^{\Omega t}$ and $\bar\xi = \xi (u)
    \e^{\Omega t}$. To preserve the diagonal form of the metric, we should
    put $\Omega\xi = \eta' \e^{2\gamma-2\alpha}$ (where $\alpha$ and $\gamma$
    are unperturbed), so that all perturbations are expressed in terms of
    $\eta(u)$. For the metric functions and the scalar field $\phi$ we
    obtain (omitting the factor $\e^{\Omega t}$)
\bear                                                      \label{g-deltas}
      \delta\alpha \eql \frac{1}{\Omega} \e^{2\gamma-2\alpha}
            [\eta'' + \eta' (2\gamma' -\alpha')],
\nn
      \delta\beta  \eql \frac{\beta'}{\Omega} \eta' \e^{2\gamma-2\alpha},
\nn
      \delta\gamma \eql \frac{1}{\Omega}
         [\Omega^2 \gamma + \gamma' \eta' \e^{2\gamma-2\alpha}],
\nn
      \delta\phi   \eql \frac{\phi'}{\Omega} \eta' \e^{2\gamma-2\alpha},
\ear
    where the prime denotes $d/du$.

    Let there be a static configuration with $\beta'\ne 0$ and $\phi'\ne 0$.
    Then, if we have nontrivial time-dependent perturbations under the gauge
    condition $\delta\beta=0$ (or $\delta\phi = 0$), \eqs (\ref{g-deltas})
    immediately lead to $\eta' = 0$, which means that our perturbation
    cannot be caused by a transformation like (\ref{dx^mu}), i.e., is
    physical. The same is true for any gauge of the form $f_1 \delta\beta +
    f_2 \delta\phi =0$ where $f_1$ and $f_2$ are any fixed functions of $u$,
    provided $f_1 \beta' + f_2 \phi' \ne 0$. The reason is that $\beta$ and
    $\phi$ are scalars with respect to coordinate transformations of the
    2-surfaces $(x^0,\ x^1$). Thus, choosing such gauges, we can be sure
    that the perturbations to be studied will be physical. For other gauges,
    involving $\delta\alpha$ and/or $\delta\gamma$, an additional inspection
    will be required.

    A more general approach to the problem of gauge in perturbation theory
    for \sph\ space-times can be found in Ref.\,\cite{seng}; though, in the
    present case, our explicit treatment seems more transparent.

\section{Stability analysis}

\subsection{The problem}

    We have considered our set of linear perturbation equations using
    two different systems of analytical computation, Maple and Mathematica,
    which made it possible to compare the results and to avoid errors.

    We use the Einstein conformal frame, in which the equations
    are much simpler, and the gauge
\bear                             \label{gauge}
    \delta\psi =0
\ear
    which is manifestly physical (see \sect 3) and, in addition, transforms
    to $\delta\phi =0$ in the Jordan frame. Moreover, according to \eq
    (\ref{SE}), we have the following relation between the metric
    perturbations:
\bear                                                    \label{p-harm}
    \delta\alpha = 2\delta\beta + \delta\gamma.
\ear
    Two independent components of the Einstein equations for perturbations
    in the gauge (\ref{gauge}) may be written as
\bearr 				                         \label{E_pert}
    \e^{2\gamma}R_1^0 = 2 [ \delta\dot{\beta}'
                - \beta'(\delta\dot\beta + \delta\dot\gamma) -
                        \gamma'\delta\dot\beta ] = 0,
\nnn
    \e^{2\alpha}R_2^2 = 2\beta''(2\delta\beta + \delta\gamma)
\nnn \cm \ \
    - 2\e^{2\beta+2\gamma}\delta\beta + \e^{4\beta}\delta \ddot \beta
                        - \delta\beta'' = 0
\ear
    Here primes denote derivatives with respect to $u$, the
    harmonic radial coordinate in the Einstein frame, $\alpha$, $\beta$ and
    $\gamma$ describe the background configuration and satisfy the static
    field equations. We separate the variables using the substitution
\bear
    \delta(r,t) = \delta(u) \e^{\Omega t},               \label{sepa}
\ear
    where $\delta$ is a perturbation of any variable in our problem.
    After substitution of (\ref{sepa}) into (\ref{E_pert}), $\delta\gamma$
    is expressed from the first equation, and then the second equation takes
    the form ($\e^{\Omega t}$ is omitted)
\beq                                \label{st1}
    \delta\beta'' - \Omega^2\ \delta\beta\ s^4(u) + F(u)\ \delta\beta'
                                  + G(u)\ \delta\beta=0 ,
\eeq
    where $s,\ F$, $G$ are functions of $u$
    obtained from the metric (\ref{ds-E}):
\bear
       F(u) \eql -2\beta''/\beta',
\nn
       G(u) \eql -2\beta''+ 2\beta''\gamma'/\beta' + 2\e^{2\beta+2\gamma},
\nn
       s(u) \eql \e^{\beta}.
\ear

    A few words about the boundary conditions. At spatial infinity the
    choice is evident: $\delta\beta \to 0$. At the transition sphere \Str\
    $\delta\beta$ should be finite, as well as its first two derivatives in
    $u$. This is necessary for the metric perturbations in the Jordan
    picture to be finite and smooth at \Str, which is easily checked using
    the transformation (\ref{map}), (\ref{ps-f}) and expressions in terms of
    the invariant length in the Jordan frame.

    As usual, we perform a transition from (\ref{st1}) to a
    Schr\"{o}dinger-like form of the perturbation equation:
\beq                            \label{schrod}
        d^2 y/dx^2 + [E - V(x)] y(x) =0,
\eeq
    where
\bear   			                        \label{transform}
    x \eql \frac{1}{m} \int s^2 d u,
\nn
    \delta\beta \eql \frac{y}{s} \exp \left(-\Half \int F d u\right),
\\
      V(x) \eql 2(\beta_{xx}/\beta_x)^2
                - (\beta_{xxx} + 2\beta_{xx}\gamma_x)/\beta_x
\nnn \ \
          + 3\beta_{xx} + 5\beta_x^2 - 4\beta_x\gamma_x -
                                		2 m^2 \e^{2\gamma-2\beta},
\ear
    where the subscript $x$ denotes $d/dx$ and $E = -m^2\Omega^2$. The
    notations are chosen in such a way that the potential $V(x)$ and the
    ``energy'' $E$ are dimensionless.  The asymptotic forms of $V(x)$ are
\bearr                                                      \label{pot1}
     V(x) \approx 2/x^3 \qquad
            (x\to\infty, \quad \mbox{spatial asymptotic}),
\nnn
     V(x) \approx -1/(4x^2) \quad
                (x\to 0, \quad \mbox{the sphere \Str}).
\ear
    Thus we have a quadratic potential well at \Str, which is placed at $x=0$
    by choosing the proper value of the arbitrary constant in the definition
    of $x$ in \eq (\ref{transform}).

    The boundary condition at spatial infinity ($u\to 0$, $x \to \infty$)
    is $y\to 0$ while the asymptotic form of any solution of (\ref{schrod})
    with $E < 0$ at large $|x|$ is
\beq                                                          \label{as1}
    y \approx C_1\e^{m\Omega |x|} + C_2\e^{-m\Omega |x|}, \qquad
                C_{1,2} = \const.
\eeq
    An admissible solution is the one with $C_1=0$, with
    only a decaying exponential.

    At the other asymptotic, $x\to 0$, the condition that follows from
    the above continuity requirements reads $y/\sqrt{x} < \infty$ whereas
    the solution of (\ref{schrod}) behaves as
\beq
    y \approx \sqrt{x} (C_3 + C_4 \ln x).                      \label{as0}
\eeq
    It follows that we must select the solution with $C_4 = 0$.
    In other words, our problem is to find out whether there is a solution
    to the boundary-value problem for \eq (\ref{schrod}) such that $y \to 0$
    as $x \to \infty$, $y/\sqrt{x} < \infty$ as $x\to 0$ and $E < 0$. In the
    remainder of the section we solve this problem.

\subsection{Summary of the solution}

    We begin with a proof of the fact that the Hamiltonian operator
    $H$ related to \eq (\ref{schrod}) is self-adjoint and is
    bounded from below. To this end, we use an auxiliary operator $T$ which
    has the same singularity at \Str\ as $H$. The one-sided boundedness
    indicates that the real part of the increment $\Omega$ cannot be
    infinite. A further comparison of $\,T$ and $H$ shows that
    the continuous parts of their spectra coincide and lie in the
    non-negative part of the real number axis. So, if there are any
    solutions of our boundary value problem with $E<0$, they belong to a
    discrete spectrum.

    To prove the existence of a solution with $E<0$ we use the well-known
    fact from quantum mechanics (its more general form is called the minimax
    principle) that the lower bound $\mu_0$ of the spectrum of an operator
    $A$ is the infimum of the functional
\beq                        				\label{func_0}
     (\psi, A \psi) ,
\eeq
    where the parentheses denote the scalar product (defined a bit later),
    the infimum is taken on the set of functions $\psi$ which lay in the
    definition domain of $T$, and the norm $\| \psi \| = 1$. Thus the value
    $(\psi, A \psi)$ for any specified function $\psi$ is an upper estimate
    for $\mu_0$, and if it is negative, then $\mu_0 < 0$. Functions which
    may closely resemble the unknown function that realizes the infimum can
    give values of the functional (\ref{func_0}) closest to $\mu_0$. We
    guess such a function, which shows that the ground state of $H$ lies
    below zero.  This function is a ground state of a certain operator which
    is similar to $H$ but simpler.

\subsection{The solution}

    Consider the auxiliary differential equation
\beq                                                	\label{aux1}
       -\frac{d^2}{dx^2}y(x) - \frac{1}{4x^2} y(x) =  E y(x)
\eeq
    and investigate the question of self-adjointness of the
    Schr\"{o}dinger operator
\beq                                            	\label{op1}
       T y(x) \equiv  -\frac{d^2}{dx^2}y(x)- \frac{1}{4x^2}y(x)
\eeq
    on the subset $D(T)$ of real Hilbert space $L_2([0, \infty))$ such that,
    for $y(x) \in D(T)$, (a) our boundary conditions (BCs) hold (so that,
    e.g., $|y|/\sqrt{x} < \infty$ as $x\to 0$) and (b) $T y(x) \in L_2$.
    The space $L_2([0,\infty))$ is a Hilbert space with an inner (scalar)
    product defined as the Lebesgue integral
\beq                                                       \label{sca_pr}
    (\varphi, \psi) = \int_{0}^{\infty} \varphi^* \psi\, dx,       
\eeq
    where the star stands for complex conjugation.
    $D(T)$ is dense in $L_2$ since $C^{\infty}_0 (0, \infty)\subset D(T)$,
    where $C^{\infty}_0(0, \infty)$ is the the subset of functions
    in $C^{\infty}(0, \infty)$ with a compact support separated from $0$.
    It is a dense subset in $L_2$ \cite{RS}.

    One can show that the operator (\ref{op1}) defined in this way is
    symmetric and therefore closable \cite{RS}. Obviously, the BCs of our
    Hilbert space are homogenous. The Schr\"odinger equation (\ref{aux1})
    related to the operator $T$ has the solution
\beq                                            \label{sol1}
              c_1\sqrt{x}K_0(\sqrt{-E}x) + c_2\sqrt{x}I_0(\sqrt{-E}x),
\eeq
    where $E$ is the ``energy'' corresponding to $-m^2\Omega^2$ of our
    problem, so to prove the instability we should show that there are
    ``quantum states'' with  $E < 0$; $K_0$ and $I_0$ are the zero-order
    modified Bessel functions of the first kind.  Neither of these
    functions, nor their any linear combination, satisfy our BCs.
    This means that the operator $T - E I$, $E<0$ has a bounded
    inverse operator $(T - E I)^{-1}$ with a definition domain dense in
    $L_2$. The existence of a reverse operator follows from the well-known
    alternative:  {\it under given homogenous BCs, either the differential
    equation $L[y] = g(x)$ has a uniquely defined solution $y(x)$, or the
    homogeneous equation $L[y]=0$ has a non-zero solution}. In our case,

\beq                                                           \label{res1}
        L[y] \equiv -\frac{d^2}{dx^2}y(x)- \frac{1}{4x^2}y(x) - E y(x).
\eeq
    The boundedness and the density property of the definition domain
    of $(T - E I)^{-1}$ in $L_2$ follow from studying the properties
    of solutions to the equation $L[y] = g(x)$ with nonzero $g(x) \in L_2$.
    The existence of $(T- E I)^{-1}$, $E<0$ means that the domain
    $(-\infty, 0) \subset \rho (T)$, $\rho (T)$ being the resolvent set
    of $T$.

    Considering in a similar way \eq (\ref{res1}) with $E > 0$, one
    can show that $[0, \infty)$ is a continuous spectrum.

    Thus we have shown that $T$ is a closed symmetric operator which
    contains real numbers in its resolvent set.  It satisfies the conditions
    of the second corollary of Theorem X.1 in \cite{RS}: {\it If the
    resolvent set of a closed symmetric operator contains at least one real
    number, then this operator is self-adjoint}. The self-adjointness of
    this operator was also mentioned in passing in Ref.\,\cite{R}.

    The proved properties of $T$ make it possible to use the wealth of
    results obtained in the theory of self-adjoint operators. In particular,
    we use the following two theorems:

\Theorem {Theorem 1 {\rm (Rellich \cite{R})}}
    {Let $A$ be a self-adjoint operator on $D(A)$ and $B$ a symmetric
    operator on $D(B)$, so that $D(B) \supset D(A)$ and
\[
    \|  B \psi \|  \leq a \|  \psi \|  + b \|  A \psi \| ,
\]
    $b<1$. Then the operator $A+B$ is self-adjoint and $D(A+B)=D(A)$.}

\Theorem {Theorem 2 {\rm (Kato \cite{R})}} {Let the conditions of Theorem 1
    hold and $A$ be bounded from below (or from above, or from both sides),
    then $A+B$ is bounded from below (or from above, or from both sides),
    but not necessarily with the same bound (bounds).}

    Considering $T$ as $A$ in these theorems, we can rewrite (\ref{schrod})
    as
\beq
        T y(x) + \widetilde{V}(x)y(x)= E y(x),
\eeq
    where
\beq
        \widetilde{V}(x) = V(x)+1/4x^2.
\eeq
    Since $\widetilde{V}$ is bounded (this is true since $\widetilde{V}(x)
    \to 0$ as $x \to \infty$, and $\widetilde{V}(x)$ is bounded everywhere),
    the conditions of Theorem 1 are fulfilled, and the operator
\beq
         H \equiv  -d^2 y/dx^2 + V(x) y(x)
\eeq
    connected with \eq (\ref{schrod}) is self-adjoint on $D(T)$.

    Using the spectral theorem for unbounded operators \cite{RS}, we prove
    that $T$ is non-negative and consequently is bounded from below.
    Therefore the operator $H$ is bounded from below too (Theorem 2).

    We now wish to show that the continuous spectrum of $H$ coincides with
    the continuous spectrum of $T$. We use the following theorem:

\Theorem {Theorem 3 {\rm \cite{RS}}} {Let $A$ be a self-adjoint operator and
    $C$ a symmetric operator
    such that $C(A^{n}+i)^{-1}$, $n \in \N$ is a compact operator. Then, if
    $B=A+C$ is self-adjoint on $D(A)$, $\sigma_{\rm ess}(A) =
    \sigma_{\rm ess}(B)$.\footnote
{We denote: $\sigma_{\rm disc} =$ discrete spectrum, $\sigma_{\rm
cont} =$ continuous spectrum, $\sigma_{\rm ess} =$ essential
spectrum \cite{RS}. In our case, $\sigma_{\rm ess}$ consists of
$\sigma_{\rm cont}$ and a possible limiting point of $\sigma_{\rm
disc}$.}}

    The compactness of the operator $\widetilde{V}(T+i)^{-1}$ follows from
    its integral representation:
\bearr
    (\widetilde{V}(T+i)^{-1} f )(x)=\int_{0}^{\infty} dy\, K(x,y)\, f(y),
\nnn
     \ K(x,y)=\widetilde{V}(x) G (x,y),
\ear
    where $G(x,y)$ is the Green function, core of the integral operator
    $(T+i)^{-1}$. As follows from the asymptotic properties of
    $\widetilde{V}(x)$ and $G(x,y)$, $K(x,y) \in L_2 ([0,\infty) \times
    [0,\infty) )$. So, $\widetilde{V}(T+i)^{-1}$ is a Hilbert-Schmidt
    operator and hence is compact \cite{RS}. The conditions of Theorem 3 are
    fulfilled, and $\sigma_{\rm ess}(H) = \sigma_{\rm ess}(T)=[0, \infty)$.
    Since $\sigma = \sigma_{\rm ess} \cup \sigma_{\rm disc}$ \cite{RS}, the
    remaining part of $\sigma(H)$ belongs to the discrete spectrum.

    \eq (\ref{st1}) may also be expressed in a non-Schr\"{o}dinger form
    using the new variable $w$ instead of $u$:
\beq                                                \label{rat_full_sub}
        u= \frac{1}{2h} \ln{ \sqrt{1/w+1}}
\eeq
    and further converted to a normal form:
\bearr \nq \label{rat_full}
   \biggl\{16E+
   \frac{11+32E}{2w}-\frac{64}{3(1+4w)}+\frac{1}{4w^2}+\frac{1}{4(1+w)^2}
\nnn
   -\frac{1}{6(1+w)}-\frac{32}{(1+4w)^2}\biggr\}n(w)
        + \frac{d^2 n(w)}{d w^2} =0,
\ear
where
\beq
    n(w)=\delta\beta(w)\exp\biggl(-\int{\frac{(2w-1)\, dw}
    					{2w(1+w)(1+4w)}}\biggr).
\eeq
    We cannot solve this equation, but its truncated version
\bearr \nq                                            \label{rat_trunc}
    \frac{d^2 n(w)}{dw^2}+
    \biggl\{16E + \frac{11+32E}{2w} + \frac{1}{4w^2}\biggr\}n(w)= 0
\ear
    has the solution:
\bearr                          \label{whit_sol}
   n(w) = c_1 M \biggl(\frac{(11+32E)}{16\sqrt{-E}},\ 0,\ 8\sqrt{-E}w \biggr)
\nnn \cm \ \
       + c_2 W \biggl(\frac{(11+32E)}{16\sqrt{-E}},\ 0,\ 8\sqrt{-E}w\biggr),
\ear
    where $M$ and $W$ are Whittaker functions. If the first argument,
    usually denoted as the index $k$, takes on the values $k= \ 1/2,\ 3/2, \
    \ldots, \ n+1/2$, $n\in \{0, \mathbb{N}\}$, then $M(k,\,0,\,w) \equiv
    W(k,\,0,\,w)$, and the corresponding quantity $E$ has the values
\beq
     E_k = -\frac{11}{32} - \frac{k^2}{8}
                \biggl(1-\sqrt{1+\frac{11}{2k^2}}\biggr),
\eeq
    which satisfy the equation
\beq
    \frac{(11+32E_k)}{16\sqrt{-E_k}}=k.
\eeq

    Performing the inverse coordinate transformation from the variables of
    (\ref{rat_full}) to the variables of (\ref{schrod}), we see that
    the functions $y$ corresponding to $W(k,0,8\sqrt{-E_k}w)$ belong to
    $D(T)$.

    Hence $E$ takes a discrete set of values in the interval $[-3/8 +
    \sqrt{23}/32, 0)$, and they lie in $\sigma_{\rm disc}$ of the
    Schr\"{o}dinger operator related to \eq (\ref{rat_trunc}). The latter
    operator has an infinite discrete spectrum with the limiting point $0$.
    Any other eigenvalue $E$ gives a solution related to the essential
    spectrum, or the resolvent set of the latter operator.  The solutions
    obtained make it possible to apply the minimax principle (\cite{RS},
    theorem XIII.1).  The theorem is applicable because $H$ is self-adjoint
    and bounded from below.

    We use the solution with $k = 1/2$ as a trial function to find an upper
    estimate of the lower bound of ${\sigma(H)}$:
\beq
    \mu_0' = \big(\psi(E_{1/2},x) ,\ H \psi(E_{1/2},x)\big),
\eeq
    where the parentheses denote the scalar product on $L_2$,
    i.e., Lebesgue integration, $\psi (E_{1/2},x)$ is a normalized function
    obtained from $W(1/2,\, 0,\, 8\sqrt{-E_{1/2}}w)$ by the
    substitution which transforms \eq (\ref{rat_full}) into (\ref{schrod}).
    Explicit integration in $x$ is impossible because we cannot represent
    $w(x)$ in elementary functions, but we can integrate in $w$ after
    necessary substitutions. It is convenient to represent $H$ as a sum of
    two Schr\"{o}dinger operators, where the first, $H'$, corresponds to \eq
    (\ref{rat_trunc}) and the second, $V'$, corresponds to the remaining
    part of the operator.  We obtain
\bearr
    \mu_0'=(\psi(E_{1/2},x) ,H'+V' \ \psi(E_{1/2},x))
\nnn
    = E_{1/2}+(\psi(E_{1/2},x) , V' \ \psi(E_{1/2},x))
\nnn
    = E_{1/2}+\frac{\int{W(1/2,0,8\sqrt{-E_{1/2}}w)^2 U(w)
            dw}}{\int{W(1/2,0,8\sqrt{-E_{1/2}}w)^2 r(w) dw}},
\ear
    where
\bearr
        U(w)=\frac{64}{3(1+4w)}-\frac{1}{4(1+w)^2}
\nnn \inch
            + \frac{1}{6(1+w)}+\frac{32}{(1+4w)^2},
\ear
    and $r(w)= 16(1+1/w)$. Integration gives $\mu_0' \simeq-0.039$. Since the
    essential spectrum begins from $0$, according to the minimax principle,
    $\mu_0'$ is an upper estimate of the ground state which is thus located
    below zero.

    Thus we have proved the existence of negative eigenvalues of the
    operator $H$ under physically justified BCs, and therefore there are
    exponentially growing solutions (at least one) of our perturbation
    equation (\ref{st1}).

    We have also solved our boundary-value problem (\ref{st1}) numerically,
    applying the Fortran procedure SLEIG and using the coordinate
    transformation (\ref{rat_full_sub}). We found a single discrete
    eigenvalue at $-0.048$, which fits our estimate of $\mu_0'$. The
    corresponding problem for \eq (\ref{rat_full}) is not suitable for
    using SLEIG because Fortran does not ``understand'' the BC
    $y/\sqrt{x} < \infty$.

    Recall that we have been working in the Einstein frame, so that the
    coordinates used cover only half of the wormhole, so that we should
    use two copies of this patch and verify whether the metric perturbations
    remain really smooth at the transition sphere \Str. Computation of
    the corresponding metric perturbations ($\delta \beta$, $\delta \gamma$)
    at the point where $f(\phi) = 0$ shows that their first derivatives in
    $l$ (where $l$ is the Gaussian radial coordinate in the Jordan picture,
    such that $g_{ll} = -1$) are zero, while the second derivatives are
    finite, so that the linearized gravity equations are there meaningful
    and hold. We conclude that the metric perturbations found are physical,
    and the wormholes under consideration are unstable under small
    spherically symmetric perturbations. The decay rate depends on the value
    of $m$ in (\ref{ds-E}), since $ E = -m^2 \Omega^2$. The \wh\ radius
    (understood, for simplicity, as $\sqrt{-g_{22}}$ at the transition
    sphere rather than throat radius which is smaller but generically of the
    same order) is proportional to $m$ and also depends on the integration
    constant in the solution of (\ref{ps-f}). Let us discuss the special
    case of conformal coupling. According to (\ref{con-y}),
    the \wh\ radius is $r_{\rm wh} = 2 m\sqrt{(1+y_0)/(1-y_0)}$. If we assume
    $y_0 \ll 1$, then both $r_{\rm wh}$ and the throat radius are
    approximately equal to $2m$. The characteristic time of decay, $\tau
    =1/\Omega$, is proportional to $m$ (which has the dimension of length):
\beq 							\label{tau}
               \tau \simeq m /\sqrt{\mu_0} \simeq 5 m.
\eeq
    For a wormhole radius of the order of a typical stellar size $\sim
    10^6$ km, the time $\tau$ is a few seconds (slightly greater than the
    time needed for a light signal to cover the stellar diameter).  If $y_0$
    increases under fixed wormhole radius, then $m$ decreases, so $\tau$
    decreases too. We see that, for all wormholes with a fixed radius, $\tau
    \leq 5 m$.

    Similar estimates can be obtained for other STT characterized by
    different $f(\phi)$.

\Acknow
 {The authors would like to thank  M.L. Fil'chenkov, V.D. Ivashchhuk, M.Yu.
  Konstantinov, V.N. Melnikov and A.B. Selivanov for useful discussions.
  K.B. acknowledges partial financial support from ISTC Project 1655.}


\small
\begin{thebibliography}{99}   

\bibitem{bg01}
        K.A. Bronnikov and S. Grinyok, \GC {7} 297 (2001);
        gr-qc/0201083.

\bibitem{br73}
        K.A. Bronnikov, {\it Acta Phys. Polon.\/} {\bf 4}, 251 (1973).

\bibitem{bar-vis99}
        C. Barcel\'o and M. Visser, \PLB {466} 127 (1999).

\bibitem{br02-CC}
        K.A. Bronnikov,
        \JMP {43} 6096 (2002); gr-qc/0201083.

\bibitem{star81}
    A.A. Starobinsky, {\it Pis'ma v Astron. Zh.\/} {\bf 7}, 67 (1981);
    {\it Sov. Astron. Lett.\/} {\bf 7}, 361 (1981).

\bibitem{Wagoner1970}
        R. Wagoner, {\it Phys. Rev.\/} {\bf D1}, 3209 (1970).

\bibitem{faraoni}
        V. Faraoni, E. Gunzig and P. Nardone,
        ``Conformal transformations in classical gravitational theories
        and in cosmology'',
        {\it Fundamentals of Cosmic Physics } {\bf 20}, 121 (1999).

\bibitem{bm-eric}
        K.A. Bronnikov and V.N. Melnikov,
        ``Conformal frames and D-dimensional gravity'',
        in: Proceedings of the 18th Course of the School on Cosmology and
        Gravitation: The Gravitational Constant. Generalized Gravitational
        Theories and Experiments (30 April-10 May 2003, Erice),
        Ed. G.T. Gillies, V.N. Melnikov and V. de Sabbata,
        Kluwer, Dordrecht/Boston/London, 2004, pp. 39--64;
        gr-qc/0310112.

\bibitem{fish}
        I.Z. Fisher, \JETF {18} 636 (1948); gr-qc/9911008.

\bibitem{hoh-vis}
        D. Hochberg and M. Visser,
        ``Geometric structure of the generic static traversable wormhole
        throat'', gr-qc/9704082.

\bibitem{cold}
    K.A. Bronnikov, G. Cl\`ement, C.P. Constantinidis and J.C. Fabris,
        {\it Phys. Lett. } {\bf A 243}, 121---127 (1998);
        \GC {4} 128--138 (1998).

\bibitem{picon02}
        C. Armendariz-Picon,
        \PRD {65}, 104010 (2002); gr-qc/0201027.

\bibitem{br-ustron}
        K.A. Bronnikov, {\it Acta Phys. Polon. } {\bf B32}, 357 (2001).

\bibitem{mitz}
        N.V. Mitskievich, ``Physical Fields in General Relativity'',
        Nauka, Moscow, 1970 (in Russian).

\bibitem{bbm70}
        N.M. Bocharova, K.A. Bronnikov and V.N. Melnikov,
        {\it Vestn. Mosk. Univ., Fiz. Astron. } No. 6, 706--709 (1970).

\bibitem{bek74}
        J.D. Bekenstein,
        {\it Ann. Phys. (USA)\/} {\bf 82}, 535 (1974).

\bibitem{bk78}
        K.A. Bronnikov and Yu.N. Kireyev,
        {\it Phys. Lett. } {\bf 67A}, 95 (1978).

\bibitem{bar-vis00}
        C. Barcel\'o and M. Visser, \CQG {17}, 3843 (2000); gr-qc/0003025.

\bibitem{birk}
        K.A. Bronnikov and M.A. Kovalchuk,
        {\it J. Phys. A: Math. \& Gen.\/} {\bf 13}, 187 (1980).

\bibitem{seng}
        U.H. Gerlach and U.K. Sengupta, ``Gauge-invariant
        perturbations on most general spherically symmetric
        space-times'', {\it Phys. Rev. }, {\bf D 19}, 2268 (1979).

\bibitem{RS}
        M. Reed and M. Simon,
        ``Methods of Mathematical Physics'', vol. I--IV,
        Academic Press, NY, 1975, 1978, 1979, 1980.

\bibitem{R}
        R.D. Richtmyer, ``Principles of Advanced Mathematical Physics'',
        vol. 1, Springer-Verlag, New York--Heidelberg--Berlin, 1978.


\end{thebibliography}
\end{document}